\documentclass[aps,prl, twocolumn, showpacs, superscriptaddress, reprint]{revtex4-1}
\usepackage{graphicx}
\usepackage{latexsym}
\usepackage{color}
\usepackage{epstopdf}
\begin{document}

\title{First Determination of the Weak Charge of the Proton}

\author{D.~Androic} 
\affiliation{University of Zagreb, Zagreb, HR 10002 Croatia }
\author{D.S.~Armstrong} 
\affiliation{College of William and Mary, Williamsburg, VA 23185 USA}
\author{A.~Asaturyan}
\affiliation{A.~I.~Alikhanyan National Science Laboratory (Yerevan Physics Institute), Yerevan 0036, Armenia}
\author{T.~Averett}
\affiliation{College of William and Mary, Williamsburg, VA 23185 USA}
\author{J.~Balewski}
\affiliation{Massachusetts Institute of Technology,  Cambridge, MA 02139 USA}
\author{J.~Beaufait}
\affiliation{Thomas Jefferson National Accelerator Facility, Newport News, VA 23606 USA}
\author{R.S.~Beminiwattha}
\affiliation{Ohio University, Athens, OH 45701 USA}
\author{J.~Benesch}
\affiliation{Thomas Jefferson National Accelerator Facility, Newport News, VA 23606 USA}
\author{F.~Benmokhtar}
\affiliation{Christopher Newport University, Newport News, VA 23606 USA}
\author{J.~Birchall}
\affiliation{University of Manitoba, Winnipeg, MB R3T2N2 Canada}
\author{R.D.~Carlini}
\email{corresponding author: carlini@jlab.org}
\affiliation{Thomas Jefferson National Accelerator Facility, Newport News, VA 23606 USA}
\affiliation{College of William and Mary, Williamsburg, VA 23185 USA}
\author{G.D.~Cates}
\affiliation{University of Virginia,  Charlottesville, VA 22903 USA}
\author{J.C.~Cornejo}
\affiliation{College of William and Mary, Williamsburg, VA 23185 USA}
\author{S.~Covrig}
\affiliation{Thomas Jefferson National Accelerator Facility, Newport News, VA 23606 USA}
\author{M.M.~Dalton}
\affiliation{University of Virginia,  Charlottesville, VA 22903 USA}
\author{C.A.~Davis}
\affiliation{TRIUMF, Vancouver, BC V6T2A3 Canada}
\author{W.~Deconinck}
\affiliation{College of William and Mary, Williamsburg, VA 23185 USA}
\author{J.~Diefenbach}
\affiliation{Hampton University, Hampton, VA 23668 USA}
\author{J.F.~Dowd}
\affiliation{College of William and Mary, Williamsburg, VA 23185 USA}
\author{J.A.~Dunne}
\affiliation{Mississippi State University,  Mississippi State, MS 39762  USA}
\author{D.~Dutta}
\affiliation{Mississippi State University,  Mississippi State, MS 39762  USA}
\author{W.S.~Duvall}
\affiliation{Virginia Polytechnic Institute \& State University, Blacksburg, VA 24061 USA}
\author{M.~Elaasar}
\affiliation{Southern University at New Orleans, New Orleans, LA 70126 USA}
\author{W.R.~Falk}
\affiliation{University of Manitoba, Winnipeg, MB R3T2N2 Canada}
\author{J.M.~Finn}
\altaffiliation{deceased}
\affiliation{College of William and Mary, Williamsburg, VA 23185 USA}
\author{T.~Forest}
\affiliation{Idaho State University, Pocatello, ID 83209 USA}
\affiliation{Louisiana Tech University, Ruston, LA 71272 USA}
\author{D.~Gaskell}
\affiliation{Thomas Jefferson National Accelerator Facility, Newport News, VA 23606 USA}
\author{M.T.W.~Gericke}
\affiliation{University of Manitoba, Winnipeg, MB R3T2N2 Canada}
\author{J.~Grames}
\affiliation{Thomas Jefferson National Accelerator Facility, Newport News, VA 23606 USA}
\author{V.M.~Gray}
\affiliation{College of William and Mary, Williamsburg, VA 23185 USA}
\author{K.~Grimm}
\affiliation{Louisiana Tech University, Ruston, LA 71272 USA}
\affiliation{College of William and Mary, Williamsburg, VA 23185 USA}
\author{F.~Guo}
\affiliation{Massachusetts Institute of Technology,  Cambridge, MA 02139 USA}
\author{J.R.~Hoskins}
\affiliation{College of William and Mary, Williamsburg, VA 23185 USA}
\author{K.~Johnston}
\affiliation{Louisiana Tech University, Ruston, LA 71272 USA}
\author{D.~Jones}
\affiliation{University of Virginia,  Charlottesville, VA 22903 USA}
\author{M.~Jones}
\affiliation{Thomas Jefferson National Accelerator Facility, Newport News, VA 23606 USA}
\author{R.~Jones}
\affiliation{University of Connecticut,  Storrs-Mansfield, CT 06269 USA}
\author{M.~Kargiantoulakis}
\affiliation{University of Virginia,  Charlottesville, VA 22903 USA}
\author{P.M.~King}
\affiliation{Ohio University, Athens, OH 45701 USA}
\author{E.~Korkmaz}
\affiliation{University of Northern British Columbia, Prince George, BC V2N4Z9 Canada}
\author{S.~Kowalski}
\affiliation{Massachusetts Institute of Technology,  Cambridge, MA 02139 USA}
\author{J.~Leacock}
\affiliation{Virginia Polytechnic Institute \& State University, Blacksburg, VA 24061 USA}
\author{J.~Leckey}
\altaffiliation{now at Indiana University, Bloomington, Indiana 47405, USA}
\affiliation{College of William and Mary, Williamsburg, VA 23185 USA}
\author{A.R.~Lee}
\affiliation{Virginia Polytechnic Institute \& State University, Blacksburg, VA 24061 USA}
\author{J.H.~Lee}
\altaffiliation{now at Institute for Basic Science, Daejeon, South Korea}
\affiliation{Ohio University, Athens, OH 45701 USA}
\affiliation{College of William and Mary, Williamsburg, VA 23185 USA}
\author{L.~Lee}
\affiliation{TRIUMF, Vancouver, BC V6T2A3 Canada}
\affiliation{University of Manitoba, Winnipeg, MB R3T2N2 Canada}
\author{S.~MacEwan}
\affiliation{University of Manitoba, Winnipeg, MB R3T2N2 Canada}
\author{D.~Mack}
\affiliation{Thomas Jefferson National Accelerator Facility, Newport News, VA 23606 USA}
\author{J.A.~Magee}
\affiliation{College of William and Mary, Williamsburg, VA 23185 USA}
\author{R.~Mahurin}
\affiliation{University of Manitoba, Winnipeg, MB R3T2N2 Canada}
\author{J.~Mammei}
\altaffiliation{now at University of Manitoba, Winnipeg, MB R3T2N2 Canada}
\affiliation{Virginia Polytechnic Institute \& State University, Blacksburg, VA 24061 USA}
\author{J.W.~Martin}
\affiliation{University of Winnipeg, Winnipeg, MB R3B2E9 Canada}
\author{M.J.~McHugh}
\affiliation{George Washington University, Washington, DC 20052 USA}
\author{D.~Meekins}
\affiliation{Thomas Jefferson National Accelerator Facility, Newport News, VA 23606 USA}
\author{J.~Mei}
\affiliation{Thomas Jefferson National Accelerator Facility, Newport News, VA 23606 USA}
\author{R.~Michaels}
\affiliation{Thomas Jefferson National Accelerator Facility, Newport News, VA 23606 USA}
\author{A.~Micherdzinska}
\affiliation{George Washington University, Washington, DC 20052 USA}
\author{A.~Mkrtchyan}
\affiliation{A.~I.~Alikhanyan National Science Laboratory (Yerevan Physics Institute),
Yerevan 0036, Armenia}
\author{H.~Mkrtchyan}
\affiliation{A.~I.~Alikhanyan National Science Laboratory (Yerevan Physics Institute),
Yerevan 0036, Armenia}
\author{N.~Morgan}
\affiliation{Virginia Polytechnic Institute \& State University, Blacksburg, VA 24061 USA}
\author{K.E.~Myers}
\altaffiliation{now at Rutgers, the State University of New Jersey, Piscataway, NJ 08854 USA}
\affiliation{George Washington University, Washington, DC 20052 USA}
\author{A.~Narayan}
\affiliation{Mississippi State University,  Mississippi State, MS 39762  USA}
\author{L.Z.~Ndukum}
\affiliation{Mississippi State University,  Mississippi State, MS 39762  USA}
\author{V.~Nelyubin}
\affiliation{University of Virginia,  Charlottesville, VA 22903 USA}
\author{Nuruzzaman}
\affiliation{Hampton University, Hampton, VA 23668 USA}
\affiliation{Mississippi State University,  Mississippi State, MS 39762  USA}
\author{W.T.H van Oers}
\affiliation{TRIUMF, Vancouver, BC V6T2A3 Canada}
\affiliation{University of Manitoba, Winnipeg, MB R3T2N2 Canada}
\author{A.K.~Opper}
\affiliation{George Washington University, Washington, DC 20052 USA}
\author{S.A.~Page}
\affiliation{University of Manitoba, Winnipeg, MB R3T2N2 Canada}
\author{J.~Pan}
\affiliation{University of Manitoba, Winnipeg, MB R3T2N2 Canada}
\author{K.D.~Paschke}
\affiliation{University of Virginia,  Charlottesville, VA 22903 USA}
\author{S.K.~Phillips}
\affiliation{University of New Hampshire, Durham, NH 03824 USA}
\author{M.L.~Pitt}
\affiliation{Virginia Polytechnic Institute \& State University, Blacksburg, VA 24061 USA}
\author{M.~Poelker}
\affiliation{Thomas Jefferson National Accelerator Facility, Newport News, VA 23606 USA}
\author{J.F.~Rajotte}
\affiliation{Massachusetts Institute of Technology,  Cambridge, MA 02139 USA}
\author{W.D.~Ramsay}
\affiliation{TRIUMF, Vancouver, BC V6T2A3 Canada}
\affiliation{University of Manitoba, Winnipeg, MB R3T2N2 Canada}
\author{J.~Roche}
\affiliation{Ohio University, Athens, OH 45701 USA}
\author{B.~Sawatzky}
\affiliation{Thomas Jefferson National Accelerator Facility, Newport News, VA 23606 USA}
\author{T.~Seva}
\affiliation{University of Zagreb, Zagreb, HR 10002 Croatia }
\author{M.H.~Shabestari}
\affiliation{Mississippi State University,  Mississippi State, MS 39762  USA}
\author{R.~Silwal}
\affiliation{University of Virginia,  Charlottesville, VA 22903 USA}
\author{N.~Simicevic}
\affiliation{Louisiana Tech University, Ruston, LA 71272 USA}
\author{G.R.~Smith}
\affiliation{Thomas Jefferson National Accelerator Facility, Newport News, VA 23606 USA}
\author{P.~Solvignon}
\affiliation{Thomas Jefferson National Accelerator Facility, Newport News, VA 23606 USA}
\author{D.T.~Spayde}
\affiliation{Hendrix College, Conway, AR 72032 USA}
\author{A.~Subedi}
\affiliation{Mississippi State University,  Mississippi State, MS 39762  USA}
\author{R.~Subedi}
\affiliation{George Washington University, Washington, DC 20052 USA}
\author{R.~Suleiman}
\affiliation{Thomas Jefferson National Accelerator Facility, Newport News, VA 23606 USA}
\author{V.~Tadevosyan}
\affiliation{A.~I.~Alikhanyan National Science Laboratory (Yerevan Physics Institute),
Yerevan 0036, Armenia}
\author{W.A.~Tobias}
\affiliation{University of Virginia,  Charlottesville, VA 22903 USA}
\author{V.~Tvaskis}
\affiliation{University of Winnipeg, Winnipeg, MB R3B2E9 Canada}
\affiliation{University of Manitoba, Winnipeg, MB R3T2N2 Canada}
\author{B.~Waidyawansa}
\affiliation{Ohio University, Athens, OH 45701 USA}
\author{P.~Wang}
\affiliation{University of Manitoba, Winnipeg, MB R3T2N2 Canada}
\author{S.P.~Wells}
\affiliation{Louisiana Tech University, Ruston, LA 71272 USA}
\author{S.A.~Wood}
\affiliation{Thomas Jefferson National Accelerator Facility, Newport News, VA 23606 USA}
\author{S.~Yang}
\affiliation{College of William and Mary, Williamsburg, VA 23185 USA}
\author{R.D.~Young}
\affiliation{University of Adelaide,  Adelaide, SA 5005 Australia}
\author{S.~Zhamkochyan}
\affiliation{A.~I.~Alikhanyan National Science Laboratory (Yerevan Physics Institute),
Yerevan 0036, Armenia}

\collaboration{The Q$_{\rm weak}$ Collaboration }
\noaffiliation

\date{\today}

\begin{abstract}
 The Q$_{\rm weak}$ experiment has measured the parity-violating asymmetry in $\vec{\rm e}$p elastic scattering at $Q^{2}$ = 0.025 $({\rm GeV/c})^{2}$, employing 145~$\mu$A of 89\% longitudinally polarized electrons on a 34.4~cm long liquid hydrogen target at 
Jefferson Lab. The results of the experiment's commissioning run, constituting approximately 4\% of the data collected in the experiment, are reported here. From these initial results the measured  asymmetry is $A_{ep} = -279 \pm 35$  (statistics) $\pm$ 31 (systematics) ppb, 
which is the smallest and most precise asymmetry ever measured in $\vec{\rm e}$p scattering. The small $Q^2$ of this experiment has made possible the first  determination of the weak charge of the proton, $Q^{p}_W$, by incorporating earlier parity-violating electron scattering (PVES) data at higher $Q^2$ to constrain hadronic corrections. The value of $Q^{p}_W$ obtained in this way 
is $Q^{p}_W ({\rm PVES})$ = $0.064 \pm 0.012$, in good agreement with the Standard Model prediction of $Q^{p}_W({\rm SM})$ = $0.0710 \pm{0.0007}$. When this result is further combined with the Cs atomic parity violation (APV) measurement, significant constraints on the weak charges of the up and down quarks can also be extracted. That PVES $+$ APV analysis reveals the neutron's weak charge to be $Q^{n}_W ({\rm PVES\!+\!APV})$ = $-0.975 \pm 0.010$.
\end{abstract}


\vspace*{0.5cm}
\maketitle

The Standard Model (SM) of electroweak physics is thought to be an effective low-energy theory of a more fundamental underlying structure. The weak charge of the proton, $Q^{p}_{W}$, is the neutral current analog to the proton's electric charge. It is both precisely predicted and suppressed in the SM and thus a good candidate for an indirect search~\cite{Musolf,Erler1,Running1,Running,SMQweak} for new parity-violating (PV) physics between electrons and light quarks. In particular, the measurement of  $Q^{p}_{W}=-2(2C_{1u} + C_{1d})$ determines~\cite{Erler1,2007C1paper} the axial electron, vector quark weak coupling constants  $C_{1i} = 2 g^e_A g^i_V$.  This information is complementary to that obtained in atomic parity violation (APV)  experiments~\cite{APV,APV1,APV2}, in particular on $^{133}$Cs where $Q_{W} (^{133}$Cs$) $=$ 55 Q^{p}_{W} + 78 Q^{n}_{W}$,  which is proportional  to a different combination,  $C_{1u} + 1.12 C_{1d}$.

The uncertainty of the asymmetry reported here is less than those of previous parity-violating electron scattering (PVES) experiments \cite{SAMPLE, SAMPLEbkwrd,   Happex1, Happex1p1, Happex1p2, Happex2He, Happex3, G0forward, G0backward, PVA41, PVA42, PVA43} directed at obtaining hadronic axial and strange form factor information~\cite{2006C1paper}. The theoretical interpretability of the Q$_{\rm weak}$ measurement is very clean as it relies primarily on those previous PVES data instead of theoretical calculations to account for residual hadronic structure effects, which are significantly suppressed at the kinematics of this experiment. 

The asymmetry $A_{ep}$ measures the cross section ($\sigma$) difference between elastic scattering of longitudinally polarized electrons with positive and negative helicity from unpolarized protons:
\begin{equation}
 A_{ep} = {{\sigma_{+}-\sigma_{-}} \over {\sigma_{+}+\sigma_{-}}} .
 \label{arc}   
\end{equation}
Expressed in terms of Sachs electromagnetic (EM) form factors~\cite{KellyFFs}
$G^{\gamma}_{\scriptscriptstyle{E}}, G^{\gamma}_{\scriptscriptstyle{M}}$, 
weak neutral form factors 
$  G^{Z}_{\scriptscriptstyle{E}},G^{Z}_{\scriptscriptstyle{M}}$ 
and the neutral weak axial form factor $G^Z_A$, the tree level asymmetry has the form~\cite{Musolf,DSARDM}: 
$$ A_{ep} = \left[- G_F Q^2 \over 4 \pi \alpha \sqrt{2}\right] \times $$
\vspace*{-0.3cm}
\begin{equation}
\left[
\varepsilon G^{\gamma}_{\scriptscriptstyle{E}} G^{Z}_{\scriptscriptstyle{E}} 
+ \tau G^{\gamma}_{\scriptscriptstyle{M}}G^{Z}_{\scriptscriptstyle{M}} 
- (1-4 \sin^2 \theta_W ) \varepsilon^{\prime} G^{\gamma}_{\scriptscriptstyle{M}} G^{Z}_{\scriptscriptstyle{A}}
\over 
\varepsilon (G^{\gamma}_{\scriptscriptstyle{E}})^{2} + \tau (G^{\gamma}_{\scriptscriptstyle{M}})^{2}
\right]
 \label{arc2}     
\end{equation}
where
\begin{equation}
\varepsilon = {1 \over 1 + 2(1 + \tau)\tan^2{\theta \over 2}}, \; \;
\varepsilon^{\prime} = \sqrt{\tau (1+\tau) (1- \varepsilon^2)}
\label{arc3}    
\end{equation}

\noindent are kinematic quantities, $G_F$ the Fermi constant, $\sin^2 \theta_W$ the weak mixing angle,
 $-Q^2$ is the four-momentum transfer squared, $\tau = Q^{2}/4M^{2}$ 
where $M$ is the proton mass,  and $\theta$ is the laboratory 
electron scattering angle. 
Eq.~\ref{arc2} can be recast as \cite{SMQweak}
\begin{equation}
{A_{ep}/A_0}  =
   Q_{W}^{p} + Q^2 B (Q^{2},\theta), \; \; \;  A_0 = \left[- G_F Q^2  \over 4 \pi \alpha \sqrt{2} \right]. 
\label{BTermEq}
\end{equation}
The dominant energy-dependent radiative correction~\cite{WWZZ} to Eq.~\ref{BTermEq} that contributes to PVES in the forward limit is the $\gamma$-Z box diagram arising from the axial-vector coupling at the electron vertex, $\Box^V_{\gamma {\rm Z}}({ E,Q^2})$.  This correction is applied directly to data  used in the $Q_{W}^{p}$ extraction prior to the fitting procedure (described below).
Then $Q^{p}_{W}$ is the intercept of  $A_{ep}/A_0$ {\em vs.} $Q^2$ in Eq.~\ref{BTermEq}. The term $Q^2 B(Q^2 , \theta)$ which contains only the nucleon structure defined in terms of EM, strange and weak form factors, is determined experimentally from existing  PVES data at higher $Q^{2}$, and is suppressed at low $Q^2$. The $Q^2$ of the measurement reported here is 4 times smaller than any previously reported $\vec{e}$p PV experiment, which ensures a reliable extrapolation to $Q^2$=0 using Eq.~\ref{BTermEq}. 

The $\gamma$-Z box diagram $\Box^V_{\gamma {\rm Z}}({ E,Q^2})$ has been evaluated using dispersion relations in ~\cite{Gorchtein1,Gorchtein2, Wally1, Wally4, Wally2, Rislow}. Interest in refining these calculations and improving their precision remains high in the theory community. Recently  Hall {\em et al.}~\cite{WallyNew}  made use of parton distribution functions to constrain the model dependence of the $\gamma$-Z interference structure functions. Combined with important confirmation from  recent Jefferson Lab (JLab) PV $\vec{\rm e}$d scattering data~\cite{Zheng}, these constrained structure functions result in the most precise calculation of $\Box^V_{\gamma {\rm Z}}$ to date. Their computed value of the contribution to the asymmetry at the Q$_{\rm weak}$ experiment's kinematics is equivalent to a shift in the proton's  weak charge of $0.00560 \pm 0.00036$, or $7.8 \pm 0.5\%$ of the SM value $0.0710 \pm 0.0007$ for $Q_W^p$~\cite{PDG2012}.  While the resulting shift in the asymmetry compared to the $Q_W^p$ term is significant, the additional $ 0.5\%$ error contribution from this correction is small with respect to our measurement uncertainty.
Charge symmetry violations are expected~\cite{Miller,Pollock,Opper,Kubis:2006cy} to be $\leq$1\% at reasonably small $Q^2$, and any remnant effects further suppressed by absorption into the experimentally-constrained $B(Q^2,\theta)$. Other theoretical uncertainties are negligible with respect to experimental errors~\cite{Running,WallyNew}.

\begin{figure}[tttht]
\hspace*{-7 mm}
{\resizebox{3.5in}{2.5in}{\includegraphics{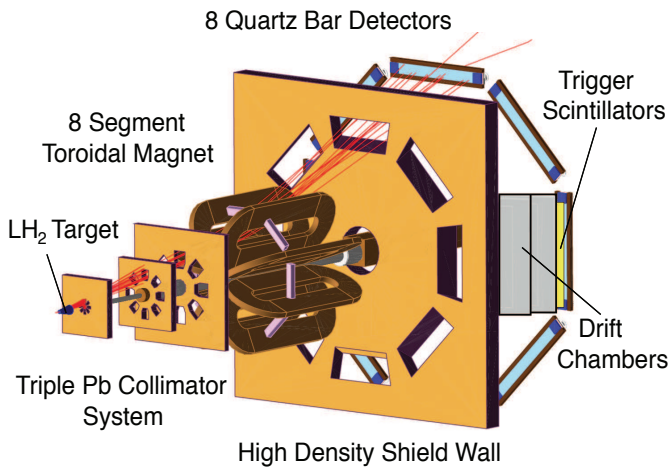}}}
\caption{The basic experimental design showing the target, collimation, magnet coils, electron trajectories, and detectors. Elastically scattered electrons (red tracks) focus at the detectors while inelastically scattered electrons (not shown), are swept away  from the detectors (to larger radii). 
The distance along the beamline from the target center to the center of the quartz bar detector array is 12.2 m.
\label{schematic-3}}
\end{figure}

The Q$_{\rm weak}$ experiment~\cite{proposal} was performed with a custom apparatus (see Fig.~\ref{schematic-3}) in JLab's Hall C.
The acceptance-averaged energy of the 145 $\mu$A, 89\% longitudinally polarized electron beam was 1.155 $\pm$ 0.003 GeV at the target center. The effective scattering angle of the experiment was 7.9$^\circ$ with an acceptance width of $\sim \pm $ 3$^\circ$. The azimuthal angle $\phi$ covered 49\% of 2$\pi$, resulting in a solid angle of 43 msr. The acceptance-averaged $Q^2$ was 0.0250 $\pm$ 0.0006 (GeV/c)$^2$, determined by simulation.


The electron beam was longitudinally polarized and reversed at a rate of 960~Hz in a pseudorandom sequence of ``helicity quartets" $(+--+)$ or $(-++-)$. The quartet pattern minimized noise due to slow linear drifts, while the rapid helicity reversal limited noise due to fluctuations in the target density and in beam properties. A half-wave plate in the laser optics of the polarized source~\cite{PolSource,QweakGun} was inserted or removed about every 8 hours to reverse the beam polarity with respect to the rapid-reversal control signals. The beam current was measured using radio-frequency resonant cavities. Five beam position monitors (BPMs) upstream of the target were used to derive the beam position and angle at the target. Energy changes were measured using another BPM at a dispersive locus in the beam line.

The intrinsic beam diameter of $\sim$250 $\mu$m  was rastered to a uniform area of $3.5 \times 3.5$ mm$^2$ at the target. 
The 57  liter, 20.00 K, liquid hydrogen target~\cite{targetPDR,targetPAVI11} consisted of a recirculating loop driven by a centrifugal pump, a 3 kW resistive heater, and a 3 kW hybrid heat exchanger making use of both 14K and 4K helium coolant. The beam interaction region consisted of a conical aluminum cell 34.4 cm long designed using computational fluid dynamics to minimize density variations due to the high power beam. The 145 $\mu$A beam deposited  1.73 kW in the target, making this the world's highest power LH$_2$ target. The measured contribution of target density fluctuations to the asymmetry width was only 37 $\pm$ 5 ppm, negligible when added in quadrature to the $\sim$250 ppm from counting statistics and other noise.

The acceptance of the experiment was defined by three Pb collimators, each with 8 sculpted openings. A symmetric array of 4 luminosity monitors was placed on the upstream face of the defining (middle) collimator~\cite{Leacock}.

A toroidal resistive DC magnet centered 6.5 m downstream of the target center consisted of 8 coils arrayed azimuthally about the beam axis. To avoid magnetic material in the vicinity of the magnet, the magnet's coil holders and support structure were composed 
of aluminum with silicon-bronze fasteners. The magnet provided 0.89 T-m at its nominal setting of 8900 A.

The magnet focused elastically-scattered electrons onto eight radiation-hard synthetic fused quartz (Spectrosil 2000) 
$\check{\text{C}}$erenkov detectors arrayed symmetrically about the beam axis 5.7~m downstream of the magnet center, and 3.3~m from the beam axis~\cite{Peiqing}. Azimuthal symmetry was a crucial aspect of the experiment's design, minimizing systematic errors 
from helicity-correlated changes in the beam trajectory  and contamination from residual transverse asymmetries.
Each detector comprised two rectangular bars 100 cm x 18 cm x 1.25 cm thick glued together into 2 m long bars. 
$\check{\text{C}}$erenkov light from the bars was read out by 12.7 cm diameter low-gain photomultiplier tubes (PMTs) through 18 cm long quartz light guides on each end of the bar assembly. The detectors were equipped with 2 cm thick Pb pre-radiators that amplified the electron signal and suppressed soft backgrounds. The detector region was heavily shielded. The beamline inside this detector hut was surrounded with 10~cm of Pb. 

With scattered electron rates of 640 MHz per detector, current-mode readout was required. The anode current from each PMT was converted to a voltage using a custom low-noise preamplifier and digitized with an 18 bit, 500 kHz sampling ADC whose outputs were integrated every millisecond. A separate PMT base was used to read out the detectors in counting (individual pulse) mode at much lower beam currents (0.1 - 200 nA) during calibration runs. During these runs, the response of each detector was measured using a system of drift chambers~\cite{Leckey} and trigger scintillators~\cite{Myers} positioned in front of two detectors at a time and removed during the main measurement. 


The raw asymmetry $A_{\rm{\it raw}}$ was calculated over each helicity quartet from the PMT integrated 
charge normalized to beam charge $Y_{\pm}$ as $A_{raw} = (Y_+ - Y_-) / (Y_+ + Y_-)$ and averaged over all detectors. Over the reported data set, $A_{\rm{\it raw}} = -169 \pm 31$~parts per billion (ppb). $A_{\rm{\it raw}}$ was corrected for false asymmetries arising from the 
measured effects of helicity-correlated beam properties to form the measured asymmetry $A_{\rm{\it msr}}$:
\begin{eqnarray}
A_{\rm{\it msr}} & = & A_{\rm{\it raw}} + A_T + A_L - \sum\limits^5_{i=1} \left( \frac{\partial A}{\partial \chi_i} \right) \Delta \chi_i \\
& = & A_{\rm{\it raw}} + A_T + A_L + A_{reg} . \label{eqn:amsr}
\end{eqnarray}
$A_T =  0 \pm 4$~ppb accounts for transverse polarization in the nominally longitudinally polarized beam~\cite{Buddhini}, and is highly 
suppressed due to the azimuthal symmetry of the experiment. It was determined from dedicated
measurements with the beam fully polarized vertically and horizontally. $A_L = 0 \pm 3$~ppb accounts for potential non-linearity in the PMT response. The $\Delta \chi_i$ are the helicity-correlated differences in beam trajectory or energy over the helicity quartet.  The slopes $\partial A / \partial \chi_i$ were determined in 6 minute intervals from linear regression using the natural motion of the beam 
and applied at the helicity quartet level. Regression corrections were studied by using different BPMs, including or excluding beam charge asymmetry (which was actively minimized with a feedback loop), and studying the effect of the corrections on the tails of the $\Delta \chi_i$ distributions. The regression correction was $A_{reg} = -35 \pm 11$~ppb. The resulting regressed asymmetry is $A_{\rm{\it msr}} = -204 \pm 31 \mbox{ ppb (statistics)} \pm 13 \mbox{ ppb (systematics)}$.

The fully corrected asymmetry is obtained from Eq.~\ref{eqn:acor} by accounting for EM radiative corrections, kinematics normalization, polarization, and backgrounds. 
\begin{equation}
{\rm A_{\it ep}} = R_{tot}\frac{ A_{\rm{\it msr}} / P  - \sum\limits^4_{i=1} f_i A_i }{1 - \sum f_i } . \label{eqn:acor}
\end{equation}
Here $R_{tot} =  R_{RC}R_{Det}R_{Bin}R_{Q^2}$, $R_{RC} =1.010 \pm 0.005$ is a radiative correction deduced from simulations with and without bremsstrahlung, using methods described in Refs. \cite{Happex1,HappexIPRC}. $R_{Det} = 0.987 \pm 0.007$ accounts for the measured light variation and non-uniform $Q^2$ distribution across the detector bars. $R_{Bin}=0.980 \pm 0.010$ is an effective kinematics correction \cite{HappexIPRC} that corrects the asymmetry from $\langle A(Q^2 )\rangle$ to $A(\langle Q^2 \rangle)$, and $R_{Q^2} = 1.000 \pm 0.030$ represents the precision in calibrating the central $Q^2$. $P = 0.890 \pm 0.018$ is the longitudinal polarization of the beam, determined using M{\o}ller polarimetry \cite{HallCMoller}. For each of the four backgrounds $b_i$, $f_i$ is the dilution (the fraction of total signal due to background "i") and $A_i$ the asymmetry.  The dilution due to all backgrounds is $f_{tot} = \sum f_i = 3.6\%$.  The statistical error in A$_{\rm{\it ep}}$ is taken as the statistical error in $A_{\rm{\it msr}}$ scaled by $\kappa = (R_{tot}/P)/(1-f_{tot}) = 1.139$.  

The largest background correction comes from the aluminum windows of the target cell ($b_1$). The cell window asymmetry was measured in dedicated runs with dummy targets and the dilution $f_1 = 3.2 \pm 0.2\%$ was obtained from radiatively-corrected measurements with the target cell evacuated. Another correction accounts for scattering sources in the beam line ($b_2$), with an asymmetry measured, along with its $f_2 = 0.2 \pm 0.1\%$ dilution, by blocking two of the eight openings in the first of the three Pb collimators with 5.1 cm of tungsten. The asymmetry measured in the detectors associated with the blocked octants was correlated to that of several background detectors located outside the acceptance of the main detectors for scaling during the primary measurement, assuming a constant dilution. The uncertainty of that correlation dominates the systematic error contribution from $b_2$.  A further correction was applied to include soft neutral backgrounds ($b_3$) not accounted for in the blocked octant studies, arising from secondary interactions of scattered electrons in the collimators and magnet. Although the corresponding asymmetry was taken as zero, an uncertainty of 100\% of the $ep$ elastic asymmetry was assigned. This dilution of $f_3 = 0.2 \pm 0.2\%$ was obtained by subtracting the blocked octant background from the total neutral background measured by the main detector after vetoing charged particles using thin scintillators. A final correction was made to account for inelastic background ($b_4$) arising from the $N \rightarrow \Delta(1232)$ transition. Its asymmetry was explicitly measured at lower spectrometer magnetic fields, and the dilution $f_4 = 0.02 \pm 0.02\%$ was estimated from simulations. 

All corrections and contributions to the systematic error in $A_{\rm{\it ep}}$ are listed in Table 1. The corrections due to multiplicative factors in $\kappa$ applied to $A_{\rm{\it raw}}$ are listed, along with the properly-normalized additive terms as defined in Eqs.~\ref{eqn:amsr} and~\ref{eqn:acor}.  The fully corrected asymmetry~\cite{Rakitha} is A$_{\rm{\it ep}} = - 279 \pm 35~\mbox{(statistics)} \pm 31~\mbox{(systematics)}$~ppb.

\begin{table}[hbp]
\centering
\begin{tabular}{l|r|r|r}

& \multicolumn{1}{c|}{Correction} & \multicolumn{2}{c}{Contribution} \\
 & \multicolumn{1}{c|}{Value (ppb)} & \multicolumn{2}{c}{ to $\Delta A_{\rm{\it ep}}$ (ppb)} \\ \hline \hline
\multicolumn{4}{c}{ Normalization Factors Applied to $A_{Raw}$ } \\ \hline
Beam Polarization $1/P$  & -21 &  \multicolumn{2}{r}{  5~ }  \\ \hline
Kinematics $R_{tot}$ & 5 & \multicolumn{2}{r}{ 9~ }\\ \hline
Bckgrnd Dilution $1/(1-f_{tot})$ & -7 & \multicolumn{2}{r}{-~ ~~} \\ \hline \hline
\multicolumn{4}{c}{ Asymmetry corrections } \\ \hline
Beam Asymmetries $\kappa A_{reg}$ &  -40~ &  \multicolumn{2}{r}{13~ } \\ \hline
Transverse Polarization $\kappa A_T$ & 0~   & \multicolumn{2}{r}{5~ } \\ \hline
Detector Linearity $\kappa A_L$ &   0~  & \multicolumn{2}{r}{4~ } \\ \hline \hline
Backgrounds& \multicolumn{1}{c|}{$\kappa P f_i A_i$} & \multicolumn{1}{c|}{$\delta(f_i)$} & \multicolumn{1}{c}{$\delta(A_i)$} \\ \hline
\hspace{0.2cm}Target Windows ($b_1$) & -58~  &  \hspace{0.4cm}  4~  &  \hspace{0.4cm} 8~  \\ \hline
\hspace{0.2cm}Beamline Scattering ($b_2$) &  11~  & 3~  & 23~  \\ \hline
\hspace{0.2cm}Other Neutral bkg ($b_3$) &  0~   &  1~  & $<1$~  \\ \hline
\hspace{0.2cm}Inelastics ($b_4$) &   1~  &  1~  & $< 1$~  \\ \hline \hline
\end{tabular}
\caption{Summary of corrections and the associated systematic uncertainty, in parts per billion. The table shows the contributions of normalization factors on $A_{\rm{\it raw}}$, then the properly normalized contributions from other sources. 
Background correction terms listed here include only $R_{tot} f_i A_i / (1-f_{tot})$; uncertainties in $A_{\rm{\it ep}}$ due to dilution fraction
and background asymmetry uncertainties are noted separately.
}
\label{tab:errorsummary}
\end{table}


Following the procedure outlined in~\cite{2006C1paper, 2007C1paper}, a global fit of asymmetries measured in PVES~\cite{SAMPLE, SAMPLEbkwrd, Happex1, Happex1p1, Happex1p2, Happex2He, Happex3, G0forward, G0backward, PVA41, PVA42, PVA43} on hydrogen, deuterium, and $^4$He targets was used to extract  $Q^{p}_{W}$ from Eq.~\ref{BTermEq}.  For this fit, EM form factors from \cite{KellyFFs} were used. The fit has effectively 5 free parameters: the weak charges C$_{1u}$ and C$_{1d}$, the strange charge radius $\rho_s$ and magnetic moment $\mu_s$, and the isovector axial form factor $G^{Z \;(T=1)}_{A}$. The value and uncertainty of 
the isoscalar axial form-factor 
$G^{Z \;(T=0)}_{A}$ (which vanishes at tree level) is constrained by the calculation of~\cite{Zhu}.
The strange quark form factors $G_E^s $=$ \rho_s Q^2 G_{D}$ and $G_M^s $=$ \mu_s G_{D}$ as well as $G^{Z \;(T=1)}_{A}$ employ a conventional dipole form~\cite{Doi} $G_{D}=(1+Q^2/\lambda^2)^{-2}$ with $\lambda$=1 $\mbox{(GeV/c)}^2$ in order to make use of PVES data up to $Q^2$=0.63 $({\rm GeV/c})^{2}$. These 4 form-factors ($G_{E,M}^s , G^{Z \;(T=0,1)}_{A}$) have little influence on the results extracted at threshold. The values for $\rho_s$ and $\mu_s$ obtained in the fit are consistent with an earlier determination~\cite{2006C1paper}  but with uncertainties $\sim 4$ times smaller.

\begin{figure}[bth]
 \hspace*{-1.5 mm}
{\resizebox{3.4in}{2.5in}{\includegraphics{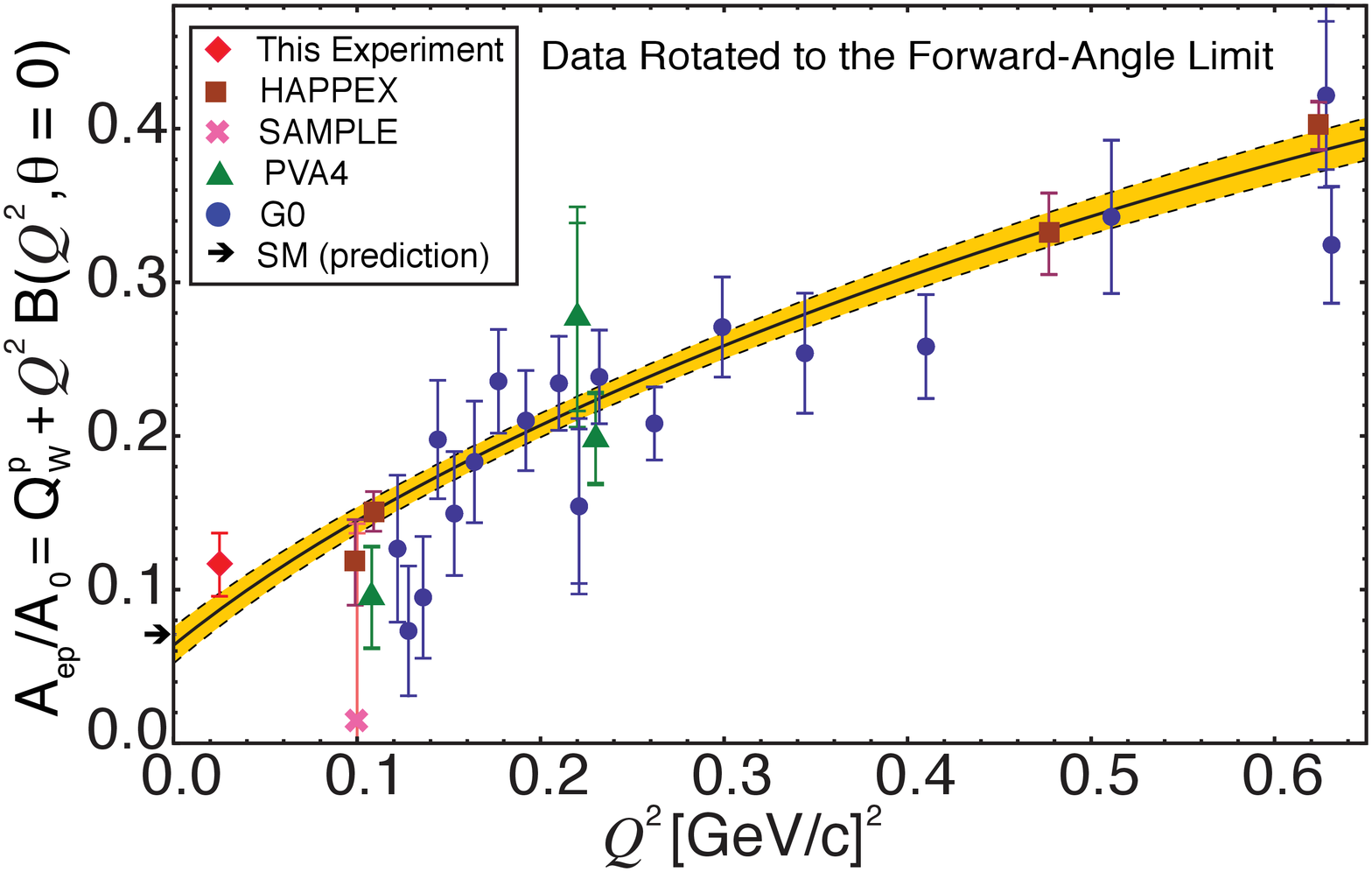}}}
 \caption{\label{BTerm}   
Global fit result (solid line) presented in the forward angle limit as reduced asymmetries derived from this measurement as well as other PVES experiments up to $Q^2 = 0.63$ $(GeV/c)^2$, including  proton, helium and deuterium data. The additional uncertainty arising from this rotation is indicated by  outer error bars on each point. The yellow shaded region indicates the uncertainty in the fit. $Q^{p}_{W}$ is the intercept of the fit. The SM prediction~\cite{PDG2012} is also shown (arrow). }
 \end{figure}

All of the $\vec{e}$p data used in the fit and shown in Fig.~\ref{BTerm} were individually corrected for the small energy dependence of the $\gamma$-Z box diagram calculated in Ref.~\cite{WallyNew}. The even smaller additional correction for the $Q^2$ dependence of  the $\gamma$-Z box diagram above $Q^2$=$ 0.025 \;  {\rm (GeV/c)} ^2$ was included using the prescription provided in Ref.~\cite{Gorchtein2} with EM form factors from Ref.~\cite{KellyFFs}. The small energy and $Q^2$ dependent uncertainties associated with the predicted corrections were folded into the systematic error of each point. The effect of either doubling, or not including the nominally forward angle $\gamma$-Z radiative correction for the 6 larger angle data $> 21^\circ$ used in the fit resulted in a change in $Q^{p}_W ({\rm PVES})< \pm$ 0.0006. 

The effects of varying the maximum $Q^2$ or $\theta$ of the data included in the fit  were studied and found to be small for data above $Q^2$$\sim$0.25 $({\rm GeV/c})^{2}$. Truncating the data set at lower $Q^2$ values tends to destabilize the fit, and enhances the sensitivity to the underlying statistical fluctuations in the data set, as reported in~\cite{2006C1paper}.
The effect of varying the dipole mass in the strange and axial form factors was also studied and found to be small, with a variation of  $< \pm$ 0.001 in $Q_W^p$ for 0.7 $({\rm GeV/c})^{2}$  $< \lambda^2 <$ 2 $({\rm GeV/c})^{2}$. Smaller values of $\lambda$ are disfavored by lattice QCD calculations of strange form factors~\cite{Doi}, and the results quickly plateau for larger values.


In order to illustrate the 2-dimensional global fit ($\theta, Q^2$) in a single dimension ($Q^2$), the angle dependence of the strange and axial form-factor contributions was removed  by subtracting
$\left[ A_{calc}(\theta, Q^2) - A_{calc}(0^\circ, Q^2) \right]$ from the measured asymmetries $A_{ep}(\theta,Q^2)$, where the calculated asymmetries $A_{calc}$ are determined from Eq.~\ref{arc2} using the results of the fit. The reduced asymmetries from this forward angle rotation of all the $\vec{e}p$ PVES data used in the global fit  are shown in Fig.~\ref{BTerm} along with the result of the
fit. The intercept of the fit at $Q^2=0$ is $Q^{p}_W ({\rm PVES})$=$0.064 \pm 0.012$. 

\begin{figure}[hbt]
{\resizebox{3.3in}{3.0in}{\includegraphics{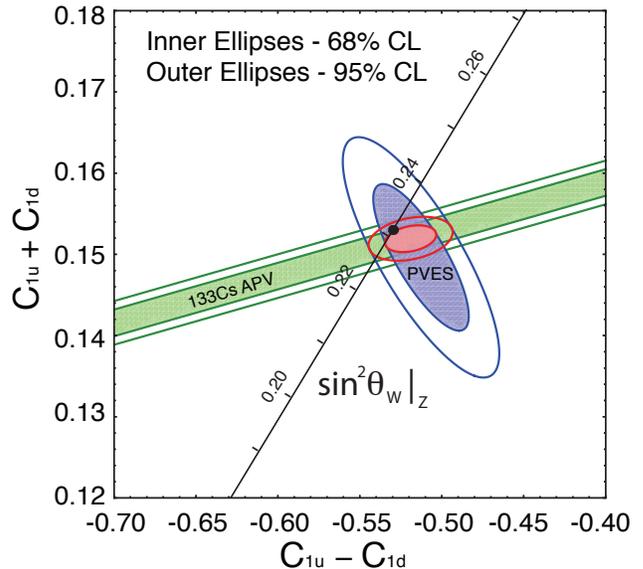}}}
 \caption{\label{C1plot}   The constraints on the neutral-weak quark coupling constants {$C_{1u}- C_{1d}$} (isovector) and 
{$C_{1u} + C_{1d}$} (isoscalar). The more horizontal (green) APV band (shown at $\Delta \chi^2 = 2.3$) provides a tight constraint on the isoscalar combination from $^{133}$Cs data. The more vertical (blue) ellipse represents the global fit of the existing $Q^2 < 0.63$ PVES data including the new result reported here at $Q^{2}$=0.025 $({\rm GeV/c})^{2}$. The smaller (red) ellipse near the center of the figure shows the result obtained by combining the APV and PVES information. The SM prediction~\cite{PDG2012} as a function of $\sin^{2}\theta_W $ in the $\overline{MS}$
scheme is plotted (diagonal black line) with the SM best fit value indicated by the (black) point at $\sin^{2}\theta_W $=0.23116. 
}
 \end{figure}

The present measurement also constrains the neutral-weak quark couplings. 
The result of a fit combining the most recent correction~\cite{Flambaum2012} to the  $^{133}$Cs APV result~\cite{APV1}, with the world PVES data (including the present measurement) is shown in  Fig.~\ref{C1plot}. 


The neutral weak couplings determined from this combined fit are $C_{1u}$=$-0.1835 \pm 0.0054$ and $ C_{1d}$=$0.3355 \pm 0.0050$, with a correlation coefficient -0.980. The couplings can be used in turn to obtain a value for $Q^{p}_{W}$, $Q^{p}_W({\rm PVES\!+\!APV})$ = $ -2(2C_{1u} + C_{1d})$=0.063$ \pm 0.012$, virtually identical with the result obtained from the PVES results alone. In addition the $C_1$'s can be combined to extract the neutron's weak charge $Q^{n}_W({\rm PVES\!+\!APV})$=$ -2(C_{1u} + 2C_{1d})$=$-0.975 \pm 0.010$. 
Both $Q^{p}_{W}$ and $Q^{n}_{W}$ are in agreement with the SM values~\cite{PDG2012} $Q^{p}_W({\rm SM})= 0.0710 \pm 0.0007$ and $Q^{n}_W({\rm SM})= -0.9890 \pm 0.0007$.

Prescriptions for determining the mass reach implied by this result can be found in the literature~\cite{Erler1,2007C1paper}.
The commissioning data reported here comprise 4\% of the total data acquired during the experiment. The final result when published will benefit from an asymmetry anticipated to have an uncertainty about 5 times smaller.

\begin{acknowledgments}
This work was supported by DOE contract No. DE-AC05-06OR23177, under which Jefferson Science Associates, LLC operates Thomas Jefferson National Accelerator Facility. Construction and operating funding for the experiment was provided through  the US Department of Energy (DOE), the Natural Sciences and Engineering Research Council of Canada (NSERC), and the National Science Foundation (NSF) with university matching contributions from the College of William and Mary, Virginia Tech, George Washington University, and Louisiana Tech University. We  wish to thank the staff of JLab, TRIUMF, and Bates, as well as our undergraduate students, for their vital support during this challenging experiment. We are also indebted to J.D. Bowman, W. Melnitchouk, A.W. Thomas, P.G. Blunden, N.L. Hall, J. Erler, and M.J. Ramsey-Musolf for many useful discussions. 

\end{acknowledgments}

\end{document}